# Lattice Parameters and Bulk Modulus of $SrTi_{1-x}Mn_xO_3$ Perovskites: A Comparison of Exchange-Correlation Functionals with Experimental Validation

Miroslav Lebeda[a,b,c], Jan Drahokoupil[a,c], Stanislav Kamba[c], Šimon Svoboda[a], Vojtěch Smola[a], Bogdan Dabrowski[d], Petr Vlčák[a]

[a] *Department of Physics, Faculty of Mechanical Engineering, Czech Technical University in Prague, Technická 4, 16607 Prague 6, Czech Republic*

[b] *Department of Solid State Engineering, Faculty of Nuclear Sciences and Physical Engineering, Czech Technical University in Prague, Trojanova 339/13, 12000 Prague 2, Czech Republic*

[c] *Institute of Physics of the Czech Academy of Sciences, Na Slovance 2, 18200 Prague 8, Czech Republic*

[d] *Institute of Physics, Polish Academy of Sciences, Aleja Lotnikow 32/46, PL-02668 Warsaw, Poland*

**Corresponding author:** Miroslav Lebeda, lebedmi2@cvut.cz

## Abstract

We assessed four exchange-correlation functionals (LDA CA-PZ, GGA parametrized by PBE, PBEsol, and WC) in predicting the lattice parameters of $SrTi_{1-x}Mn_xO_3$ perovskites, assuming cubic structures. Predictions were verified using X-ray diffraction (XRD) for Mn content of $x$ = 0.0, 0.1, 0.2, 0.3, 0.5, 1.0, confirming cubic symmetry and a linear decrease in lattice parameters with increasing Mn. PBEsol, and WC demonstrated the highest precision (deviations < 0.20 %). Additionally, bulk moduli were calculated using the same functionals and verified with the experimental bulk modulus of $SrTiO_3$ (183 ± 2 GPa, Pulse-Echo method). The predicted bulk moduli exhibited a slow, linear increase with increasing Mn. The best correspondence with the experimental bulk modulus was achieved by PBEsol and WC (deviations < 0.7 %). These findings highlight the reliability of PBEsol and WC functionals for accurately modeling structural properties of $SrTi_{1-x}Mn_xO_3$ perovskites, having better precision than commonly employed LDA and PBE functionals.

**Key words:** SrTi1-xMnxO3 perovskites; Exchange-correlation functionals; Lattice parameters; Bulk modulus

## 1. Introduction

Strontium titanate ($SrTiO_3$, STO), a typical perovskite-type metal oxide, is a widely used dielectric material due to its excellent chemical and thermal stability, non-toxicity, affordability, and versatility. These properties have made $SrTiO_3$ an utilized component in diverse applications, such as microwave devices, filters, gas sensors, capacitors, superconductors, or substrates for deposition of various thin films [1–3]. Notably, $SrTiO_3$ has garnered attention as a promising photocatalyst, particularly in the water splitting process for generating clean, renewable hydrogen fuel [4]. However, its large bandgap (~ 3.2 eV) limits its light absorption to ultraviolet spectrum, which accounts only for a small fraction of sunlight, thereby restricting its effectiveness in visible-light photocatalysis [5]. To address this limitation, doping $SrTiO_3$ with transition metal ions have been subjected to research [6]. It was found out that such doping not only enhances its photocatalytic activity but also broadens its applications to include random-access memory, anode materials for lithium-ion batteries, and advanced energy storage systems.

Among the chemically modified variants of $SrTiO_3$, $SrTi_{1-x}Mn_xO_3$ solid solution has attracted significant interest due to its largely improved photocatalytic properties and high dielectric constant. The incorporation of Mn introduces new impurity energy levels, narrowing the bandgap and enhancing light absorption while increasing the material's specific surface area and reducing electron-hole recombination. As a result, $SrTi_{1-x}Mn_xO_3$ has shown great potential as a photocatalyst for the degradation of dye pollutants and as a component in advanced energy storage systems [7].

Typically, $Mn^{4+}$ substitutes $Ti^{4+}$ within $SrTiO_3$ structure due to their similarity in ionic radii ($Mn^{4+}$: 0.530 Å, $Ti^{4+}$: 0.605 Å) [8,9]. This induces important structural changes. Theoretical predictions of these changes are necessary for tailoring the material's application potential. *Ab initio* methods, particularly density functional theory (DFT), are widely used for such predictions. However, the accuracy of DFT calculations depends heavily on the chosen exchange-correlation functional, with no single functional accurately capturing properties of all material systems due to its inherent limitations and specific optimizations. Therefore, benchmarking the performance of different functionals is important for understanding their strengths, weaknesses, and applicability to specific materials [10,11].

In this work, we employed four functionals (LDA, PBE, PBEsol, WC) to evaluate their effectiveness in predicting the lattice parameters and bulk modulus of $SrTi_{1-x}Mn_xO_3$ series supposing cubic structures. The simulations were compared with room-temperature experimental lattice parameters obtained via X-ray diffraction (XRD) of ceramics for six different Mn concentrations ($x$ = 0.0, 0.1, 0.2, 0.3, 0.5, 1.0). Additionally, the computed bulk moduli were compared with experimental value for $SrTiO_3$ determined by Pulse-Echo method. Our findings indicate that PBEsol and WC provide the most accurate reproduction of lattice parameters and bulk modulus, with negligible precision difference among the two.

# 2. Experimental, Computational, and Materials Details

## 2.1 Materials

$SrTiO_3$ ceramic samples were prepared via the conventional mixed-oxide route using high-purity $SrCO_3$ and fine-grained $TiO_2$ powders, with stoichiometry adjusted to $Sr_{1.000}Ti_{1.001}O_3$. The powders were mixed in a cyclohexanolic suspension, ball-milled for 3 hours, and calcined at 1050 °C for 18 hours to ensure perovskite phase formation, verified by XRD. Cylindrical samples were cold-pressed and sintered at 1380 °C for 7 hours, achieving 98.8% theoretical density with an average grain size of 1–2 µm. Post-sintering, samples underwent polishing and chemical treatments, including etching in orthophosphoric acid. The synthesis process was carried out following the method described in [12].

$SrTi_{1-x}Mn_xO_3$ ceramic samples were synthesized using a two-step solid-state method following the similar routine as described for $Sr_{1-x}Ba_xMnO_3$ to maintain the pseudocubic perovskite structure [13]. Precursor materials were first prepared in flowing Ar gas at ~1400 °C to obtain the single-phase oxygen-reduced samples, which were then annealed in oxygen at 350 °C to achieve an oxygen stoichiometry of 3.000 ± 0.002, confirmed by the thermogravimetric analysis.

## 2.2 Experimental Characterization

### 2.2.1 XRD Settings

XRD measurements were conducted at room temperature using a PANalytical X'Pert Pro powder diffractometer with a cobalt anode ($K_{\alpha1}$ = 1.789 Å, $K_{\alpha2}$ = 1.793 Å). The Bragg-Brentano geometry was used with the following instrumental settings: a 1 ° divergence slit, 0.02 rad Soller slits, a 10 mm mask, and a β-filter. The Rietveld refinement method within TOPAS software [14] was employed to obtain lattice parameters and symmetry information.

### 2.2.2 Pulse-Echo Settings

The bulk modulus was measured using the Pulse-Echo method with a WaveRunner LT264M oscilloscope and a JSR Ultrasonics DPR300 for signal generation and reception. To enhance the transmission of longitudinal waves, samples were coated on both sides with propylene glycol. Only $SrTiO_3$ sample was suitable for this measurement, yielding reliable results that aligned with existing literature. In contrast, the rest of $SrTi_{1-x}Mn_xO_3$ samples produced unreliable bulk modulus values, likely due to inhomogeneities or porosity. As a result, these measurements were not included in this study.

## 2.3 Computational Settings

The simulations were conducted using supercells derived from the initial cubic unit cell of $SrTiO_3$ with 3.905 Å lattice parameter, having *Pm-3m* (No. 221) symmetry [15]. The original cell was expanded to form a 2 x 2 x 2 supercell (**Fig. 1**), resulting in a structure containing 8 Ti ions. These Ti ions were incrementally replaced with Mn ions at random positions to model different Mn concentrations, corresponding to $x$ = 0.000, 0.125, 0.250, 0.375, 0.500, 0.625, 0.750, 0.875, 1.000.

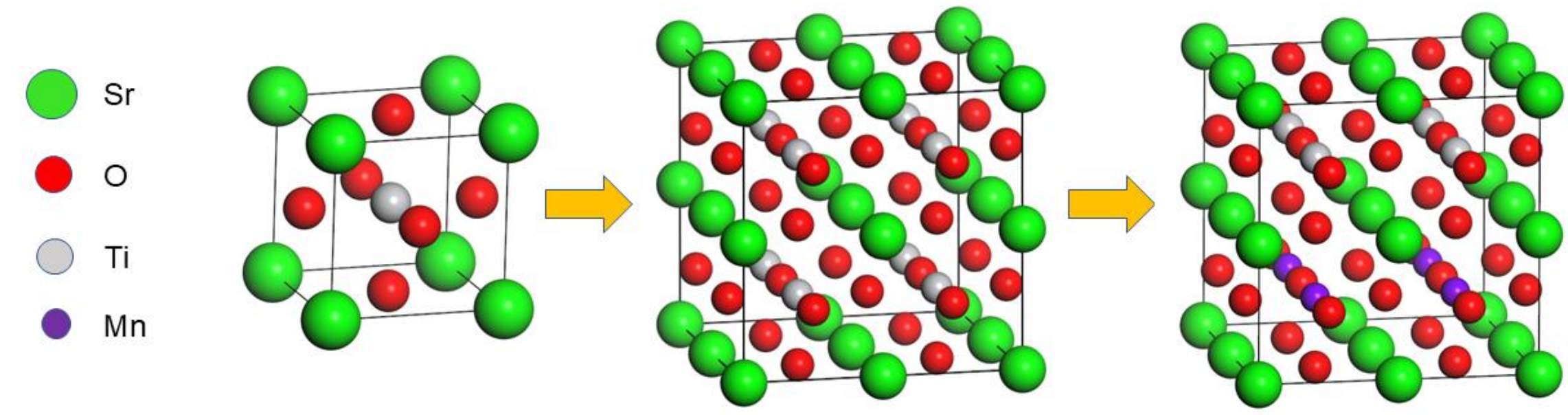

**Figure 1**: A 2 x 2 x 2 supercell of $SrTiO_3$ with four Mn ions substituting Ti sites.

DFT simulations were carried out using the CASTEP module within the Dassault Systèmes BIOVIA Materials Studio software [16]. The exchange-correlation function was set as either local density approximation (LDA) based on the Ceperley and Alder data as parameterized by Perdew and Zunger (CA-PZ), or the following generalized gradient approximation (GGA) variations: Perdew-Burke-Ernzerhof (PBE), PBE for solids (PBEsol), and Wu-Cohen (WC). Within the CASTEP calculations, the fix occupancy option was set to off to allow variable electronic occupancies during the self-consistent field (SCF) iterations.

Ultrasoft pseudopotentials were applied to simplify the electron wavefunctions near atomic nuclei. The SCF energy tolerance was set as 5 x $10^{-7}$ eV/atom. Convergence tests for cutoff energy ($E_{cut}$) and $k$-point sampling were performed on $SrTiO_3$ and $SrTi_{0.875}Mn_{0.125}O_3$ in order to balance the computational precision with the constraints of available resources. Based on these tests, an $E_{cut}$ of 380 eV was selected, as ultrasoft pseudopotentials typically require lower cutoff energy. Monkhorst-Pack $k$-point sampling was set to 4 x 4 x 4. This allowed for the precision of ~10 meV/atom. While such level of precision is less stringent than the commonly used range of 1–5 meV/atom, typically required for high-accuracy calculations such as phase transitions or defect energies, it is considered to be sufficient for the objectives of this study, which focus on identifying trends in lattice parameters and bulk moduli across different Mn concentrations and functionals. Consistent settings were applied across the modeled $SrTi_{1-x}Mn_xO_3$ series to ensure reliable trends and provide reasonable precisions.

Geometry optimization was performed on the simulated $SrTi_{1-x}Mn_xO_3$ series, under the assumption of cubic symmetry to determine the lattice parameters. The bulk moduli were calculated by examining the energy-volume relationship of the optimized unit cells and fitting the data to the Birch-Murnaghan equation of state [17]:

$$E(V) = E_0 + \frac{9V_0B_0}{16}\left\{\left[\left(\frac{V_0}{V}\right)^{\frac{2}{3}} - 1\right]^3 B_0' + \left[\left(\frac{V_0}{V}\right)^{\frac{2}{3}} - 1\right]^2 \left[6 - 4\left(\frac{V_0}{V}\right)^{\frac{2}{3}}\right]\right\},$$

where $E_0$ is the equilibrium energy at equilibrium volume $V_0$, $B_0$ is the bulk modulus at equilibrium volume $V_0$, and $B_0'$ is the pressure derivative of the bulk modulus. These four parameters were treated as fitted parameters. The fitting was performed over an 18% range around both sides of the equilibrium volume for each composition.

# 3. Results

## 3.1 XRD Characterization of $SrTi_{1-x}Mn_xO_3$

The measured X-ray diffractograms for the $SrTi_{1-x}Mn_xO_3$ series are presented in **Fig. 2**. Rietveld analysis confirms that all samples have a cubic structure at room temperature, with notable variations in lattice parameters. These lattice parameters are summarized in **Tab. 1**, revealing a distinct trend: as the concentration of Mn increases, the lattice parameter decreases approximately linearly (also visible in **Fig. 4**).

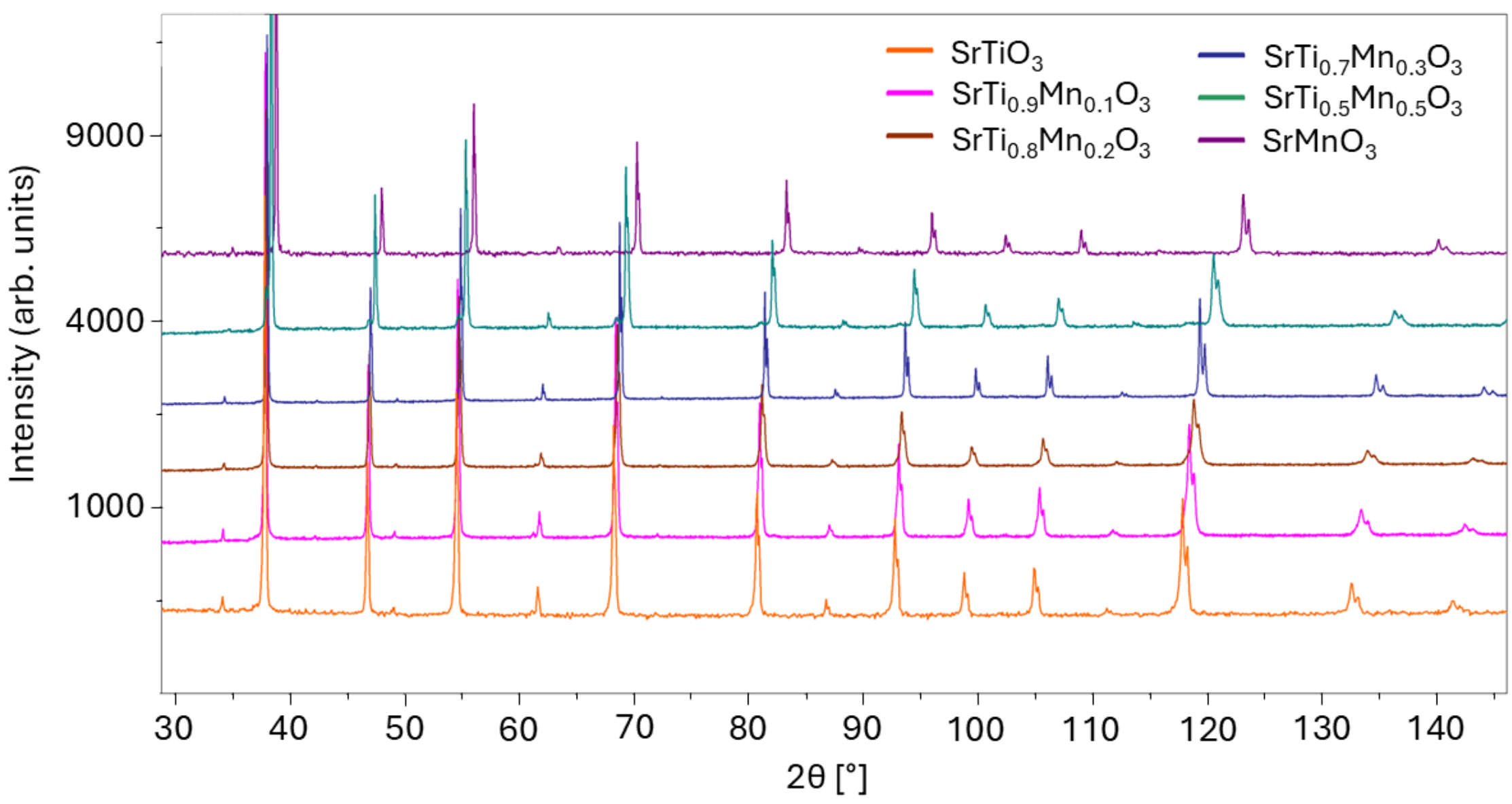

**Figure 2**: Room-temperature X-ray diffractograms of the $SrTi_{1-x}Mn_xO_3$ series. All patterns correspond to a cubic perovskite structure.

**Table 1**: Experimental lattice parameters for $SrTi_{1-x}Mn_xO_3$ determined by Rietveld analysis of X-ray diffractograms.

| Sample | Lattice parameter (Å) |
|---|---|
| $SrTiO_3$ | 3.905(9) |
| $SrTi_{0.9}Mn_{0.1}O_3$ | 3.895(9) |
| $SrTi_{0.8}Mn_{0.2}O_3$ | 3.887(4) |
| $SrTi_{0.7}Mn_{0.3}O_3$ | 3.876(0) |
| $SrTi_{0.5}Mn_{0.5}O_3$ | 3.860(2) |
| $SrMnO_3$ | 3.806(0) |

Additionally, diffraction profiles at higher angles display a left-sided shoulder (gradual onset) for the $SrTi_{0.9}Mn_{0.1}O_3$. and $SrTi_{0.8}Mn_{0.2}O_3$ samples. This behavior is marked with arrows in **Fig. 3** at 2θ angles between 118 ° and 119 °. A single cubic phase did not accurately fit the shoulders, yielding an RWP factor of 4.3 %. The shoulders can be likely attributed to either inhomogeneity with lower Mn content (higher lattice parameters) or the presence of a phase with lower tetragonal symmetry (low temperature phase of $SrTiO_3$ [18]). We have tested both possibilities. A tetragonal phase fit showed no significant improvement in modelling the shoulders, with the RWP factor marginally decreasing to 4.2 %. In contrast, a model with two cubic phases notably improved the fit (RWP = 2.4 %), with the refined lattice parameters of 3.887 Å for the major phase and 3.895 Å for the minor phase. This minor phase corresponds to a Mn content of approximately $x = 0.115$, based on the linear lattice parameter approximation for the $SrTi_{1-}$ $_x$Mn$_x$O$_3$ series. Therefore, the shoulders suggest a low amount of inhomogeneity within these samples.

Another possible explanation for the inhomogeneity could be attributed to the formation of extended defects involving coupled Mn ions: the two or more coupled Mn next neighbour ions instead of the randomly distributed isolated Mn ions. This phenomenon likely arises during the two-step synthesis process, where high-temperature reduction in an Ar atmosphere selectively reduces $Mn^{4+}$ to $Mn^{3+}$, leaving Ti ions unaffected and maintaining their 6-coordinated state. At elevated temperatures (above 800 °C) cation diffusion may cause two Mn ions to approach one another, facilitating the removal of the oxygen ion between them and forming $Mn^{3+}$ ions in pyramidal coordination. This structural defect persists upon cooling to room temperature. Subsequent low-temperature oxygenation (below 600 °C) is insufficient to significantly alter this defect structure. At higher Mn concentrations ($x > 0.2$) the increased likelihood of Mn clustering and random distribution of Ti ions within the Mn matrix reduces the degree of inhomogeneity. Exploring this hypothesis is however beyond the scope of this work.

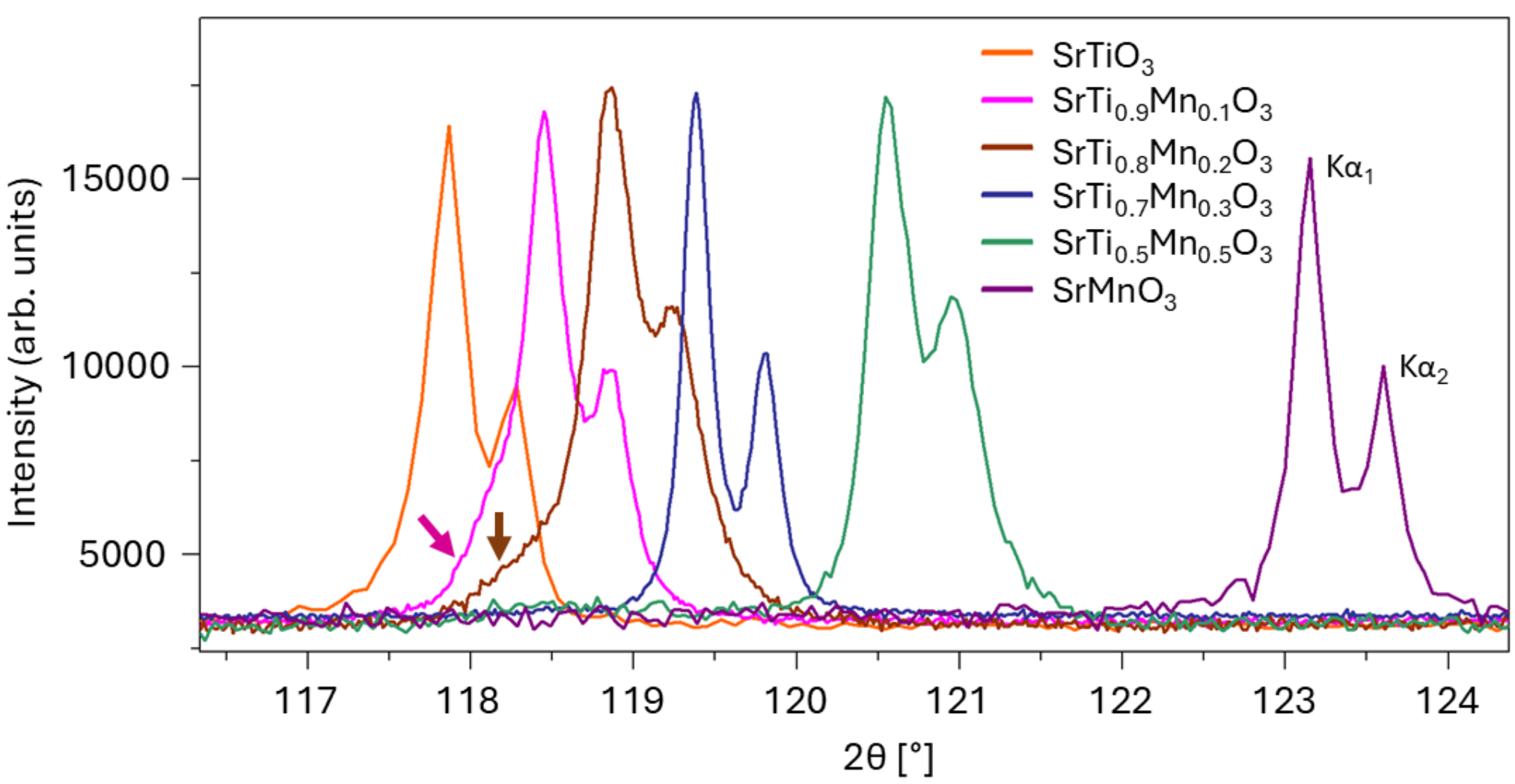

**Figure 3:** Intensity dependence on the 2θ angle for diffraction profiles around 120 ° ((123) planes) of SrTi$_{1-x}$Mn$_x$O$_3$ samples. Within this 2θ range, a clear splitting between the K$\alpha_1$ and K$\alpha_2$ components is visible. The shift of diffraction peaks toward higher angles with increasing Mn concentration reflects a decreasing lattice parameter. Arrows indicate the left-sided shoulders observed for $x$ = 0.1 and 0.2, highlighting their minor degree of inhomogeneity.

### 3.2 Effect of Exchange-Correlation Functional on $SrTi_{1-x}Mn_xO_3$ Lattice Parameters and Bulk Modulus

The choice of exchange-correlation functional significantly influences DFT simulation results. We examined the impact of LDA CA-PZ and GGA variations of PBE, PBEsol and WC functionals on the lattice parameters of $SrTi_{1-x}Mn_xO_3$ perovskites. All functionals exhibited a linear decrease in lattice parameters with increasing Mn concentration, in agreement with experimental data (**Fig. 4**). Linear approximations of these dependencies ($y$ = a$x$ + b, where $y$ represents the lattice parameter and $x$ the Mn content in $SrTi_{1-x}Mn_xO_3$) allowed for the comparison of DFT values with experimental results. The corresponding linear equations are listed in **Tab. 2**.

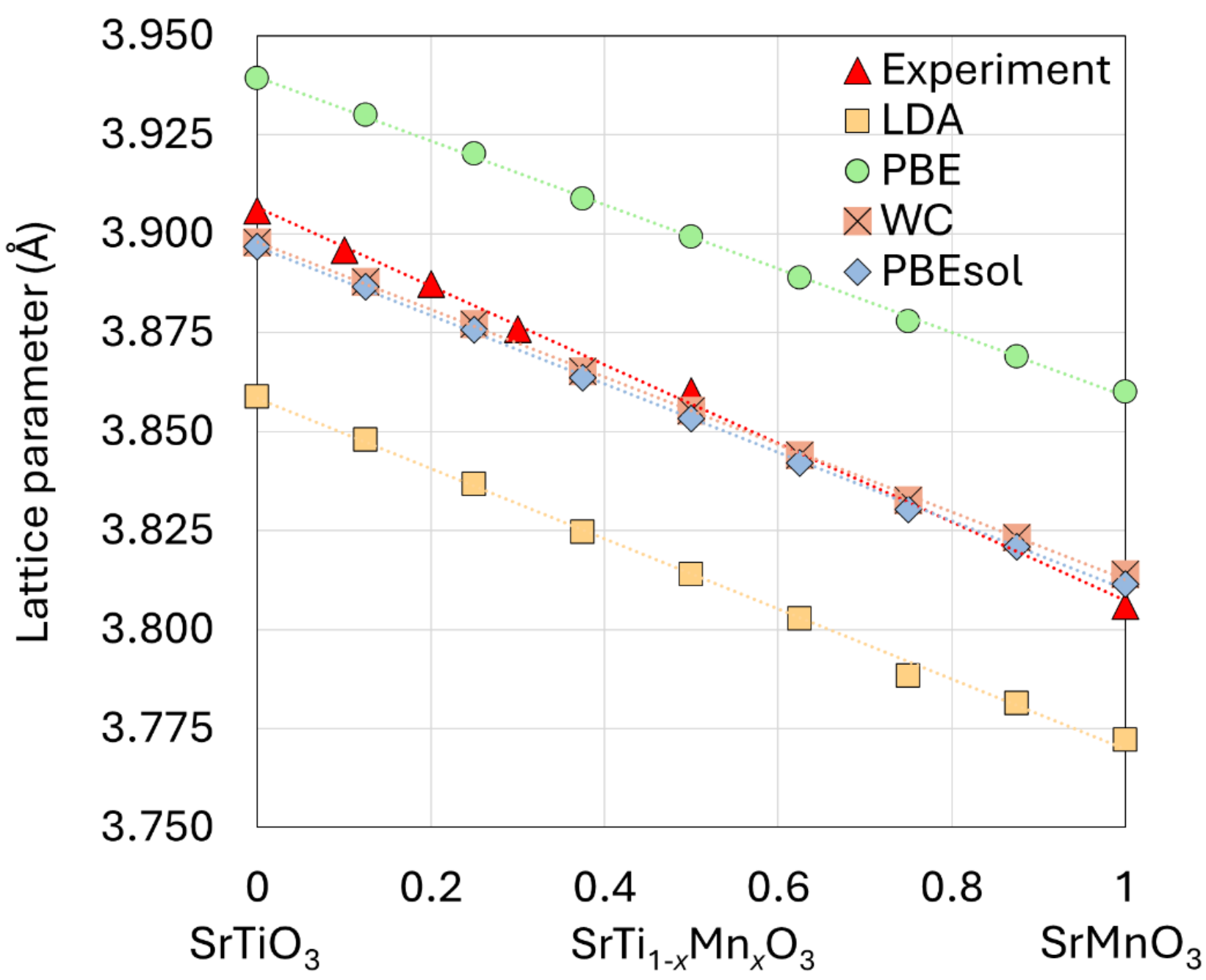

**Figure 4:** The lattice parameters for cubic $SrTi_{1-x}Mn_xO_3$ obtained via XRD and from DFT calculations using LDA CA-PZ and GGA exchange-correlation functionals parametrized by PBE, PBEsol, and WC.

**Table 2:** The linear approximations of lattice parameters for cubic $SrTi_{1-x}Mn_xO_3$ obtained by XRD and from DFT calculations using LDA CA-PZ and GGA exchange-correlation functionals parametrized by PBE, PBEsol, and WC.

| Exchange-correlation functional | Linear approximation of lattice parameter |
|---|---|
| LDA | -0.0886$x$ + 3.858 |
| PBE | -0.0806$x$ + 3.940 |
| PBEsol | -0.0867$x$ + 3.897 |
| WC | -0.0852$x$ + 3.898 |
| Experiment | -0.0993$x$ + 3.907 |

**Tab. 3** provides a detailed comparison of experimental lattice parameters with predictions from DFT calculations. The comparison is expressed in terms of relative deviation (δ) with our experimental values serving as the reference. Among the tested functionals, PBEsol and WC provided the best agreement with experiments, having minimal deviations. For PBEsol, deviations ranged from -0.20 % for $SrTiO_3$ to 0.11 % for $SrMnO_3$, while WC showed a similar performance with deviations from -0.18 % to 0.18 %. In contrast, LDA consistently underestimated the lattice parameters, with deviations around 1.2 % for most compositions, whereas PBE consistently overestimated, with deviations around 1.0 %, except for $SrMnO_3$, where the deviation increased to 1.4 %. These findings confirm that linear approximation of supercell-derived values using PBEsol and WC very well predicts experimental lattice parameters for any Mn concentration within the cubic structures of $SrTi_{1-x}Mn_xO_3$.

**Table 3:** Relative deviations in the lattice parameters of $SrTi_{1-x}Mn_xO_3$ between experimental values by XRD and theoretical predictions from DFT using LDA CA-PZ and GGA exchange-correlation functionals parametrized by PBE, PBEsol, and WC. Negative values indicate that the DFT-calculated lattice parameters are smaller than the experimental values.

| $x$ in $SrTi_{1-x}Mn_xO_3$ | LDA δ (%) | PBE δ (%) | WC δ (%) | PBEsol δ (%) |
| --- | --- | --- | --- | --- |
| 0.0 | -1.20 | 0.90 | -0.18 | -0.20 |
| 0.1 | -1.18 | 0.95 | -0.14 | -0.17 |
| 0.2 | -1.20 | 0.95 | -0.16 | -0.19 |
| 0.3 | -1.15 | 1.03 | -0.09 | -0.13 |
| 0.5 | -1.20 | 1.03 | -0.12 | -0.16 |
| 1.0 | -0.96 | 1.40 | 0.18 | 0.11 |

The bulk moduli of the $SrTi_{1-x}Mn_xO_3$ perovskites were calculated using the Birch-Murnaghan equation of state by fitting the energy-volume relationships on the supercells obtained from DFT geometry optimizations. **Fig. 5** demonstrates the quality of such the Birch-Murnaghan fit for the $SrTiO_3$ calculated using PBE. The fit aligned well with the DFT data on the compressed side of the lattice, even for volume reductions of up to 30 % relative to the equilibrium volume. For expanded volumes, the fit begun to deviate from the DFT calculations when volumes exceeded approximately 20 % above the equilibrium volume.

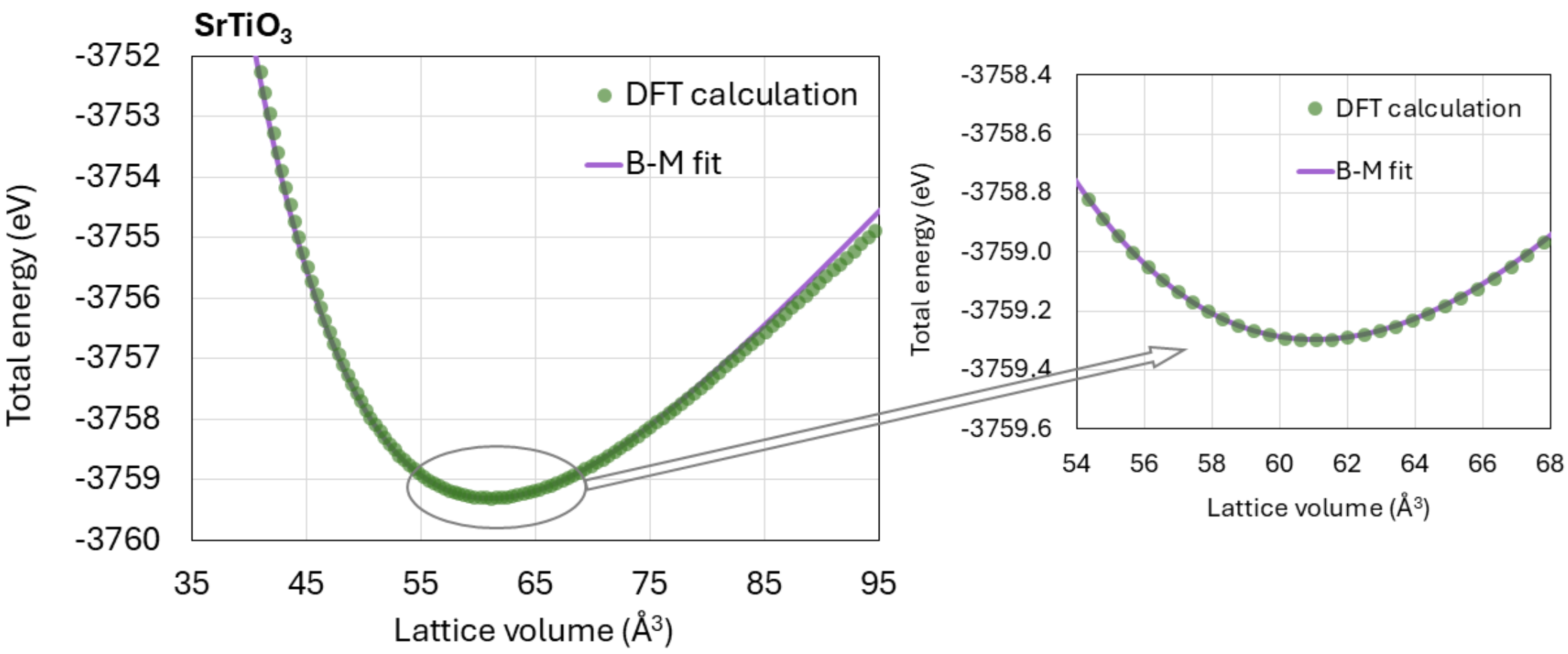

**Figure 5:** Example of a Birch-Murnaghan (B-M) fit to the $SrTiO_3$ energy-volume relationship from DFT GGA PBE calculations.

The resulted bulk moduli showed a moderately steep linear increase as the Mn concentration increases across all tested functionals (**Fig. 6**). For $SrTiO_3$, the experimental bulk modulus determined by the Pulse-Echo method was 183 ± 2 GPa, aligning well with the literature value of 174 GPa [19]. Among the tested functionals, PBEsol and WC showed excellent agreement with our experimental value of 183 GPa, with deviations of 0.6 % and 0.7 %, respectively. In contrast, the LDA functional overestimated the bulk modulus at 201 GPa, deviating by 9.7 %, while PBE underestimated it at 168 GPa, showing a deviation of 8.2 %. This behavior (overestimation by LDA and underestimation by PBE) is opposite to their trends in predicting lattice parameters. Overall, PBEsol and WC provided the best agreement with experimental data for both bulk modulus and lattice parameters.

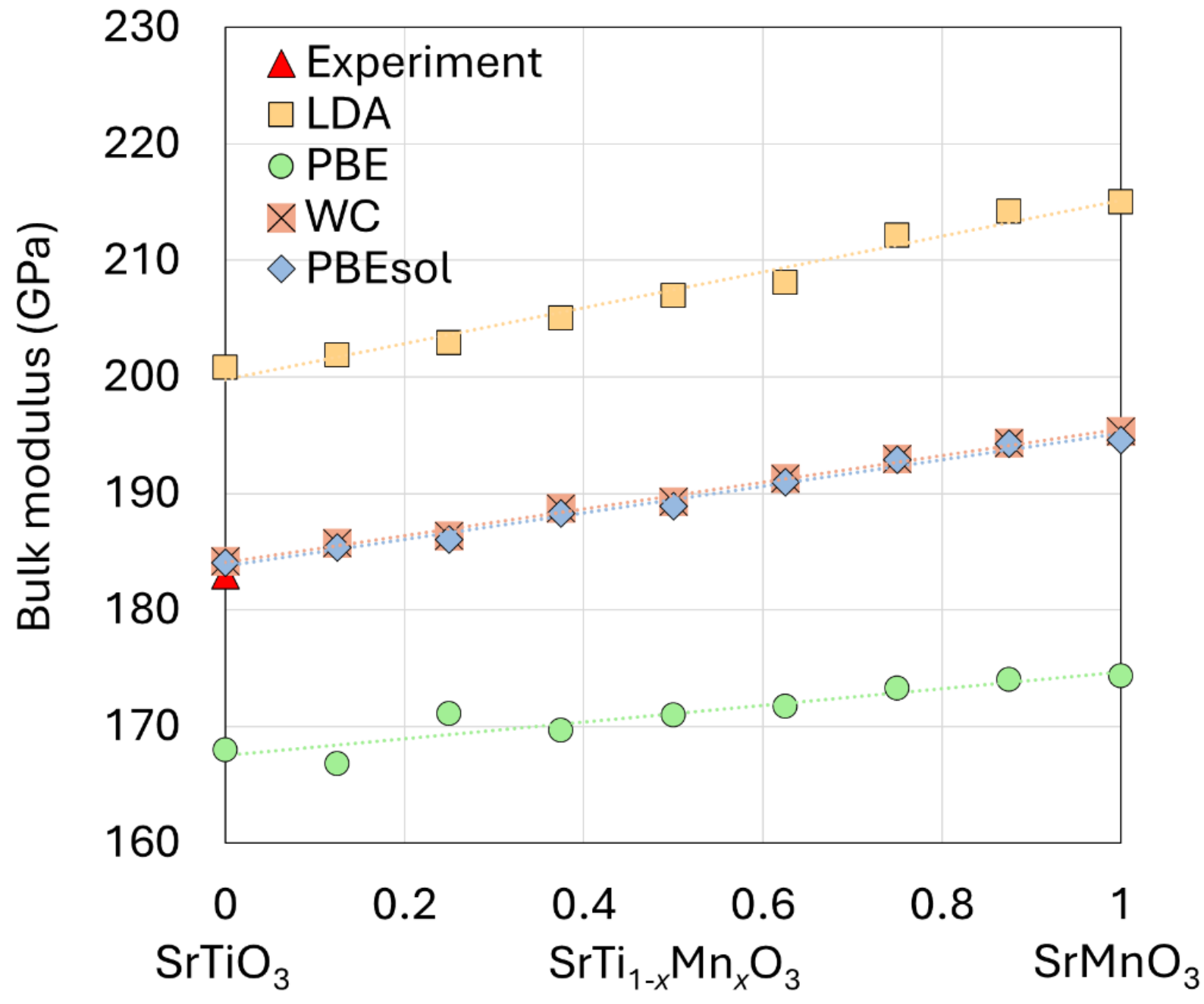

**Figure 6:** Theoretical bulk moduli for cubic $SrTi_{1-x}Mn_xO_3$ obtained from DFT calculations using LDA CA-PZ and GGA exchange-correlation functionals parametrized by PBE, PBEsol, and WC. For comparison, the experimental bulk modulus of $SrTiO_3$ determined using the Pulse-Echo method is also shown.

## 4. Conclusions

The influence of four exchange-correlation functionals (LDA CA-PZ, GGA parametrized by PBE, PBEsol, WC) on the lattice parameters of $SrTi_{1-x}Mn_xO_3$ perovskites in cubic phases was investigated and compared with experimentally determined lattice parameters obtained via XRD for compositions $x$ = 0.0, 0.1, 0.2, 0.3, 0.5, 1.0. XRD analysis confirmed that all samples have cubic symmetry, with lattice parameters decreasing linearly as Mn content increased. DFT calculations mirrored this trend, with PBEsol and WC functionals providing the best agreement to experimental values, exhibiting relative deviations of less than 0.20 % and 0.18 %, respectively, across all Mn concentration. LDA and PBE demonstrated lower precision, with deviations ranging from ~1.0 % to 1.4 %. Additionally, bulk moduli were calculated across all four functionals and compared with the experimental bulk modulus of $SrTiO_3$, determined to be 183 ± 2 GPa via the Pulse-Echo method. DFT results revealed a slow increasing linear tendency of bulk modulus with increasing Mn content. For $SrTiO_3$, PBEsol and WC again demonstrated the closest agreement with experimental data, having relative deviations of less than 0.7 %. These results indicate that among the tested functionals, PBEsol and WC are the most suitable exchange-correlations functionals for modeling the structural and elastic properties of $SrTi_{1-x}Mn_xO_3$, providing better precision than commonly used LDA or PBE functionals.

## 5. Acknowledgments

This work was supported by the Grant Agency of the Czech Technical University in Prague [grant No. SGS24/121/OHK2/3T/12] and project TERAFIT - CZ.02.01.01/00/22_008/0004594 co-financed by European Union and the Czech Ministry of Education, Youth and Sports. We would also like to express our gratitude to Michaela Janovska for helping with the Pulse-Echo measurements.